\documentclass[%
 aip,
 amsmath,amssymb,floatfix,
 reprint,%
]{revtex4-1}

\usepackage[american]{babel}

\usepackage{graphicx}
\usepackage{dcolumn}
\usepackage{bm}

\usepackage[utf8]{inputenc}
\usepackage[T1]{fontenc}
\usepackage{mathptmx}
\usepackage{etoolbox}
\usepackage{placeins}

\usepackage{siunitx}
\usepackage[version=4]{mhchem}

\usepackage{tikz}

\makeatletter
\def\@email#1#2{%
 \endgroup
 \patchcmd{\titleblock@produce}
  {\frontmatter@RRAPformat}
  {\frontmatter@RRAPformat{\produce@RRAP{*#1\href{mailto:#2}{#2}}}\frontmatter@RRAPformat}
  {}{}
}%
\makeatother
\begin{document}

\preprint{AIP/123-QED}

\title[Ferroelectric domain walls for environmental sensors]{Ferroelectric domain walls for environmental sensors}
\author{L. Richarz}
\affiliation{ 
Department of Materials Science and Engineering, NTNU Norwegian University of Science and Technology, Trondheim, Norway
}%

\author{I. C. Skogvoll}
\author{E. Y. Tokle}
\affiliation{ 
Department of Materials Science and Engineering, NTNU Norwegian University of Science and Technology, Trondheim, Norway
}%
\author{K. A. Hunnestad}
\affiliation{ 
Department of Materials Science and Engineering, NTNU Norwegian University of Science and Technology, Trondheim, Norway
}%
\affiliation{ 
Department of Electronic Systems, NTNU Norwegian University of Science and Technology, Trondheim, Norway
}%
\author{U. Ludacka}
\affiliation{ 
Department of Materials Science and Engineering, NTNU Norwegian University of Science and Technology, Trondheim, Norway
}%
\affiliation{%
Department of Physics, NTNU Norwegian University of Science and Technology, Trondheim, Norway
}%

\author{J. He}
\affiliation{ 
Department of Materials Science and Engineering, NTNU Norwegian University of Science and Technology, Trondheim, Norway
}%

\author{E. Bourret}
\affiliation{%
Materials Sciences Division, Lawrence Berkeley National Laboratory, Berkeley, CA, USA
}%

\author{Z. Yan}
\affiliation{%
Materials Sciences Division, Lawrence Berkeley National Laboratory, Berkeley, CA, USA
}%
\affiliation{%
Department of Physics, ETH Zurich, Zurich, Switzerland
}%

\author{A.T.J. van Helvoort}
\affiliation{%
Department of Physics, NTNU Norwegian University of Science and Technology, Trondheim, Norway
}%

\author{J. Schultheiß}
\affiliation{
Department of Materials Science and Engineering, NTNU Norwegian University of Science and Technology, Trondheim, Norway
}%

\author{S. M. Selbach}
\affiliation{ 
Department of Materials Science and Engineering, NTNU Norwegian University of Science and Technology, Trondheim, Norway
}%

\author{D. Meier}
\affiliation{ 
Department of Materials Science and Engineering, NTNU Norwegian University of Science and Technology, Trondheim, Norway
}%

\date{\today}

\begin{abstract}

Domain walls in ferroelectric oxides provide fertile ground for the development of next-generation nanotechnology. Examples include domain-wall-based memory, memristors, and diodes, where the unusual electronic properties and the quasi-2D nature of the walls are leveraged to emulate the behavior of electronic components at ultra-small length scales. Here, we demonstrate atmosphere-related reversible changes in the electronic conduction at neutral ferroelectric domain walls in Er(Mn,Ti)O$_3$. By exposing the system to reducing and oxidizing conditions, we drive the domain walls from insulating to conducting, and vice versa, translating the environmental changes into current signals. Density functional theory calculations show that the effect is predominately caused by charge carrier density modulations, which arise as oxygen interstitials accumulate at the domain walls. The work introduces an innovative concept for domain-wall-based environmental sensors, giving an additional dimension to the field of domain wall nanoelectronics and sensor technology in general.



\end{abstract}

\maketitle

\section{Introduction}

Environmental sensors play a key role in modern electronics and the Internet of Things, having profound implications for various fields, ranging from home electronics and autonomous transport to safeguarding the ecosystem. The sensor's task is to provide information about parameters, such as temperature, pressure, humidity, concentration of gases, and soil moisture, converting respective changes into electronic signals. One example is semiconductor-based gas sensors, which rely on conductivity variations that arise when a semiconducting oxide (e.g., \ce{SnO_2} or \ce{ZnO}) is exposed to a certain target gas.\cite{Nikolic2020} Other concepts include polymers, nanotubes, and 2D materials for sensing applications.\cite{Dai2020,Schedin2007} Although such technologies are well-established and versatile environmental sensors are commercially available, several challenges remain. Traditional semiconductor sensors work best at relatively high operating temperatures and their recovery time after the detection of an event is relatively long.\cite{Liu2012} 
In addition, there is a pressing demand for miniaturization, enabling a higher packing density, the development of multi-functional environmental sensors, as well as concepts for spatially resolved sensing at the local scale.

In this context, domain walls in ferroelectric oxides are particularly promising as a novel type of ultra-small sensing system. The domain walls represent natural interfaces with thicknesses approaching the unit cell level, and they are known to strongly interact with oxide defects\cite{Schultheiss2020, Schaab2018,Farokhipoor2011,Farokhipoor2012,Seidel2010,Guyonnet2011,Rojac2017}, which can be leveraged in sensing applications. Depending on the structure and charge state of a domain wall, it can either attract or repel oxygen defects and, thereby, amplify oxidation- and oxidation-reduction reactions\cite{Becher2015,Schaab2018,Skjærvø2018,Schultheiss2020,Campanini2020}. For example, it has been shown that oxygen vacancies have a tendency to accumulate at domain walls in BiFeO$_3$\cite{Campanini2020,Rojac2017,Bencan2020} and LiNbO$_3$\cite{Eggestad2024} and co-determine the local electronic response. In addition, domain walls can facilitate large ionic mobilities\cite{Salje2000}, further enhancing the reactivity and potentially reducing the recovery time after exposure to, e.g., oxygen-rich or -poor atmospheres. Most importantly for sensing applications, the direct relation between the concentration of oxygen defects at the domain walls and their electronic transport behavior allows to readily translate environmentally driven variations in oxygen concentration into measurable changes in the local conductivity. 

In spite of these promising physical and chemical properties, the application potential of ferroelectric domain walls in sensor technology remains largely unexplored. Up to now, most studies focused on domain wall nanoelectronics, where the walls are utilized to emulate the behavior of classical electronic components at ultra-small length scales\cite{Seidel2009,Mundy2017,Schaab2018,Scharma2017,McConville2020}, and the implementation of domain walls in neuromorphic and other unconventional computing schemes.~\cite{Everschor-Sitte2024} Readers interested in a more comprehensive coverage are referred to recent reviews\cite{Nataf2020,Meier2022,Sharma2022}. In contrast, much less is known about domain-wall-based sensing. In a recent study, the enhanced conductance at domain walls in a \ce{LiNbO_3} thin film was utilized to enhance the film's response to changes in temperature\cite{Geng2021}. Similarly, the defect concentration at the domain walls in \ce{LiNbO_3} was shown to slow down electron-hole recombination, which is of interest for light sensing and in-memory computing.~\cite{Zhang2024}


Here, we investigate the relation between reducing/oxidizing environmental conditions and the electronic transport properties at ferroelectric domain walls in Er(Mn,Ti)O$_3$. We show that reversible changes in conductance occur at neutral domain walls when exposed to reducing and oxidizing atmospheres, changing the domain walls from insulating to conducting and vice versa. Our density functional theory (DFT) calculations reveal that oxygen-defect-driven variations in the local carrier concentration, rather than band-gap-related effects, drive this change in transport behavior, providing a microscopic understanding. The results demonstrate the general possibility of leveraging ferroelectric domain walls as active units for sensing applications, translating environmental parameters into measurable electronic signals.


\section{Exposure to reducing atmosphere}

For our study, we use the ferroelectric semiconductor Er(Mn,Ti)O$_3$ with 0.2\% Ti doping (see Methods Section for more information on the crystal). For this material, the basic domain wall physics are well-understood\cite{Choi2010,Meier2012,Schultheiss2020,Wu2012,Smabraten2018,Schoenherr2019,Turner2018,Du2011,McCartan2024,Holtz2017}, and it is predicted that the system has a propensity to accumulate oxygen defects (i.e., interstitials) at its neutral domain walls\cite{Schaab2018,Schultheiss2020}. In general, the system displays an outstanding chemical flexibility and tunable semiconducting transport properties (p-type\cite{Holstad2018}), that can be controlled via the oxygen content, i.e., the oxygen off-stoichiometry in \ce{Er(Mn{,}Ti)O}$_{3+\delta}$, with typical values $\delta > 0$ in the as-grown state.\cite{Remsen2011,Danmo2024, Swierczek2017,Remsen2011,Skjærvø2016} This chemical flexibility, in combination with the affinity to accumulate oxygen defects at the neutral domain walls, makes it an ideal model system for domain-wall-based environmental sensing.

\ce{ErMnO_3} is one of the hexagonal manganites (\textit{R}MnO$_3$ with \textit{R} = Sc, Y, In, Dy to Lu), in which the spontaneous polarization appears as a secondary effect, caused by a structural instability of the paraelectric high-temperature phase ($T_C \approx 1170$~K\cite{Chae2012}) that leads to a tripling of the unit cell and the formation of a polar axis\cite{Artyukhin2014} (improper ferroelectricity\cite{Fujimura1996,VanAken2004} with a spontaneous polarization $P \approx 5.6$~\SI{}{\micro C m^2})\cite{Fujimura1996}. The structural instability drives the formation of topologically protected structural vortex lines that govern the domain formation and serve as anchor points for the ferroelectric domain walls as explained elsewhere\cite{Artyukhin2014}. As a consequence, a robust 3D network of interconnected domain walls arises that we will explore in the following for the development of ultra-small sensors.

\begin{figure}[h]
    \centering
    \includegraphics[width = 0.47\textwidth]{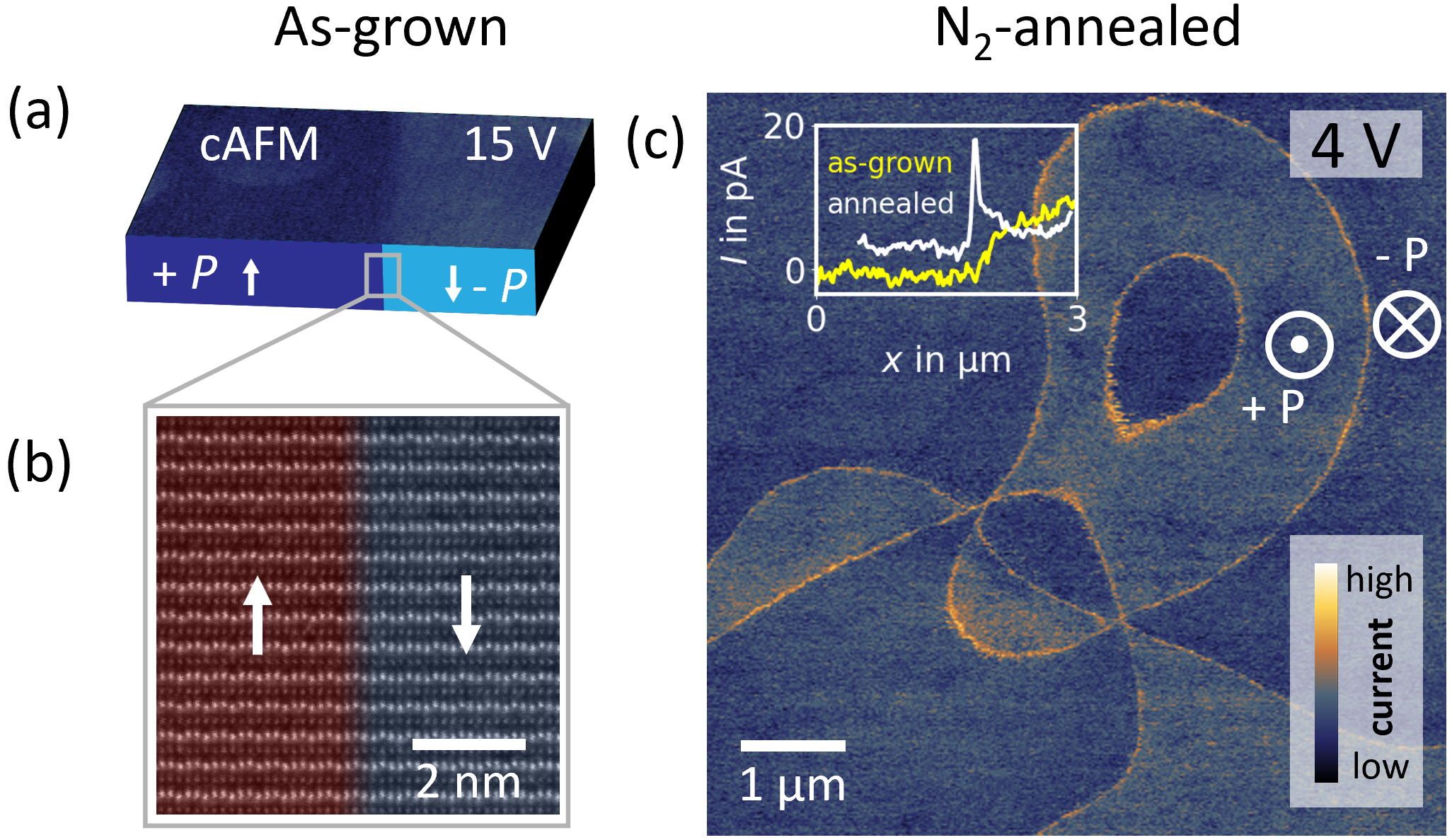}
    \caption{\textbf{Domain wall conductance and annealing in reducing atmosphere.} (a) Illustration of the transport behavior of $+P$ and $-P$ domains in \ce{Er(Mn{,}Ti)O_3} (out-of-plane polarization). The top corresponds to a cAFM map ($V_{bias}=15$~V), showing lower (dark) and higher (bright) conductance in $+P$ and $-P$ domains, respectively. Polarization directions are indicated by white arrows. (b) Representative HAADF-STEM image of a neutral domain wall with color overlay (red: $+P$, blue: $-P$). (c) cAFM scan ($V_{bias}=4$~V) obtained on the same sample as in (a) after annealing in \ce{N_2} at 300 $^\circ$C for \SI{48}{hrs}. For consistency, the length and current scales are identical in (a) and (c); cAFM scans are performed with a diamond-coated DEP01 tip. Representative line plots comparing the conductance evolution between $+P$ and $-P$ in (a) and (c) are shown in the inset of (c).}
    \label{fig:annealing}
\end{figure}

We begin by analyzing the general impact of variations in oxygen off-stoichiometry on the domain wall conductance by exposing our sample to reducing conditions. Figure~\ref{fig:annealing}(a) shows a representative conductive atomic force microscopy (cAFM) scan recorded on the [001]-surface of a \ce{Er(Mn{,}Ti)O_3} single crystal ($P$ out-of-plane) before exposure. The cAFM map is recorded with a bias voltage, $V_{bias}$, of \SI{15}{V} applied to the back-electrode while the tip is grounded (see Methods Section for further details) and shows the characteristic transport behavior of [001]-oriented hexagonal manganites.~\cite{Wu2010} Due to the screening of negative bound charges at the surface of $-P$ domains by mobile hole carriers and barrier effects at the tip-sample interface, a higher conductance (bright) is measured on $-P$ domains compared to the less conducting (dark) $+P$ domains.~\cite{Wu2010}

 \begin{figure*}[t]
	\centering
	\includegraphics[width = 0.9\textwidth]{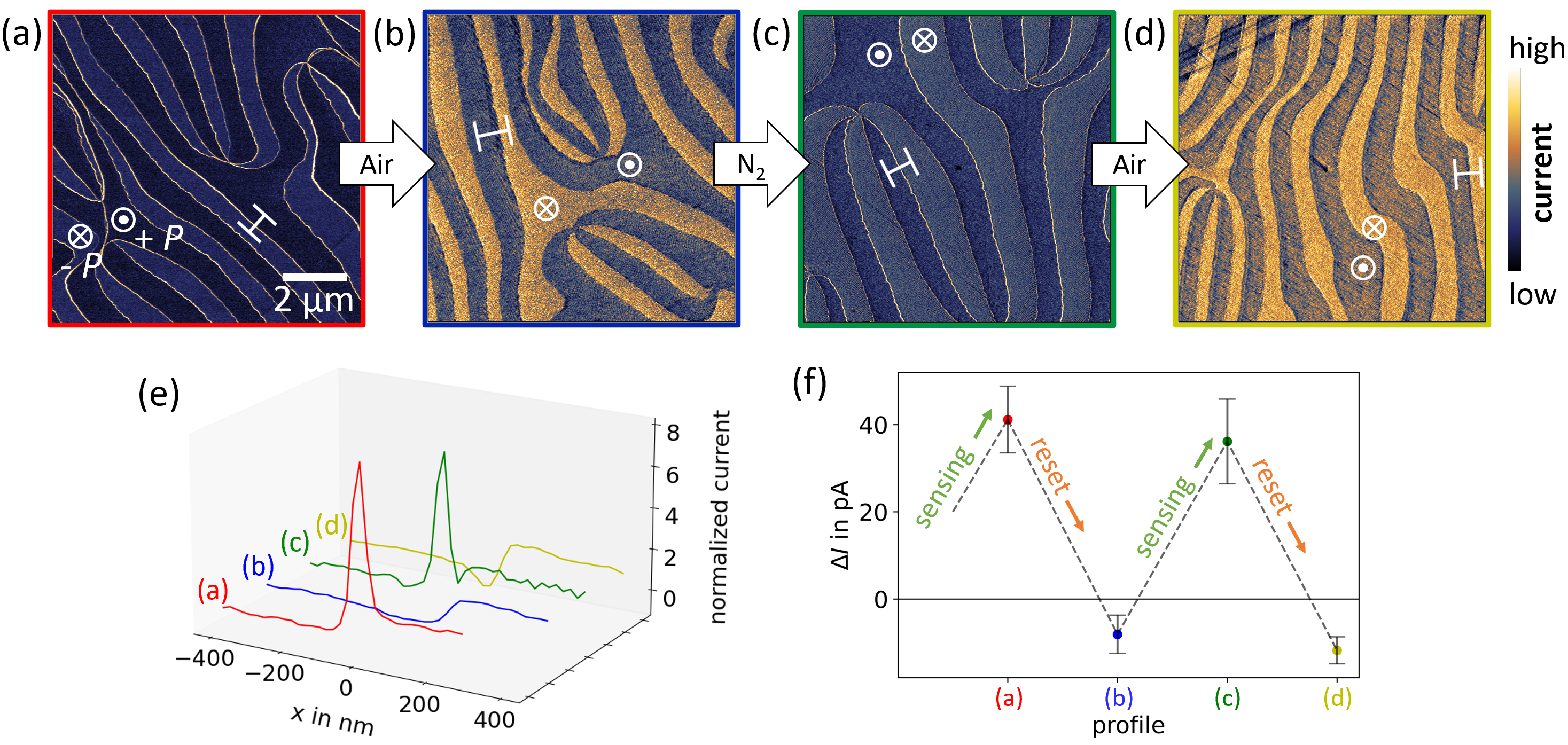}
	\caption{\textbf{Reversibility of atmosphere-driven changes in domain wall conductance.} cAFM scans taken (a) after annealing in \ce{N_2} at \SI{300}{\degreeCelsius} for \SI{48}{hrs}, (b) after subsequent heating up to 200 $^\circ$C in \ce{N_2} and synthetic air, (c) after repeated annealing in \ce{N_2} at 300 $^\circ$C for 48 hrs, and (d) after repeated heating up to 200 $^\circ$C in \ce{N_2} and synthetic air (see Supplementary Information for details). (e) Current profiles extracted along the white markings in (a)-(d). The profiles are normalized such that the $+P$ ($-P$) domains have an average current value of 0 (1). (f) Relative domain wall conductance $\Delta I$ for the four profiles in (e). By exposing the system to the different atmospheres, $\Delta I$ switches between conductive ($\Delta I \gg 0$) to insulating ($\Delta I < 0$) behavior. The cAFM scans are performed with diamond-coated tips (DEP$01$ for (a) to (c) and CDT-NCHR for (d)) and bias voltages of (a) \SI{2}{V}, (b) \SI{4}{V}, (c) \SI{5}{V} and (d) \SI{5}{V}.}
	\label{fig:revers}
\end{figure*}

The $+P$ and $-P$ domains in Figure~\ref{fig:annealing}(a) are separated by a neutral $180^\circ$ domain wall as highlighted by the representative high-angle annular dark-field scanning transmission electron microscopy (HAADF-STEM) image in Figure~\ref{fig:annealing}(b). The HAADF-STEM image is recorded viewing along the $[11\bar{0}]$ direction with bright dots indicating the positions of the \ce{Er} atoms and shows the atomic-scale structure of a neutral domain wall in Er(Mn,Ti)O$_3$~\cite{Holtz2017}. The local polarization direction can readily be determined based on the \ce{Er} displacement patterns ($+P$ = \raisebox{0.2em}{$\bullet\bullet$}$\bullet$ and $-P$ = $\bullet\bullet$\raisebox{0.2em}{$\bullet$}, where $\bullet$ indicates the position of an \ce{Er} atom) as explained in ref.~\cite{Das2014,Holtz2017}. Most importantly for this work, the cAFM data in Figure~\ref{fig:annealing}(a) shows that in the as-grown state, the transport properties at the neutral domain wall are similar to the bulk and cannot be distinguished from the surrounding domains.

Figure~\ref{fig:annealing}(c) presents a conductance map gained on the same sample with \SI{4}{V} bias voltage after exposing the sample to reducing conditions. More specifically, the cAFM image is taken after annealing in nitrogen (\ce{N_2}) at \SI{300}{\degreeCelsius} for \SI{48}{h} (heating/cooling rate: \SI{\pm 200}{\degreeCelsius/h}).
Despite the lower bias voltage compared to the pre-annealing scan (Figure~\ref{fig:annealing}(a)), comparable current values are measured for the domains as shown in the inset to Figure~\ref{fig:annealing}(c). This observation reflects a general increase of the conductance in the near-surface region probed by cAFM ($\lesssim  100$~nm\cite{Roede2022}) in response to the annealing.
The most pronounced effect, however, is a qualitative change regarding the transport behavior at the neutral domain walls. After the exposure to reducing conditions, their conductance is about three times higher than in the domains, reflecting that the neutral domain walls have a much higher sensitivity to the environmental history than the domains.

\section{Reversibility of environmentally driven effects}

In the next step, we explore to what extent the domain wall conductance can be reset to the initial state and test the repeatability of the process. For this purpose, we introduce an additional heating step at up to 200 $^\circ$C involving synthetic air (labeled "air") as summarized in Figure~\ref{fig:revers} (see Supplementary Table~S1 and Methods Section for details).

The cAFM image series in Figure~\ref{fig:revers} shows conductance maps that were taken on the same sample while going through consecutive annealing cycles. To avoid potential imprints from previous scans, the cAFM maps are recorded in different positions as visible from the different domain structures in Figure~\ref{fig:revers}(a) to (d). Figure~\ref{fig:revers}(a) shows a representative conductance map ($V_{bias}=2$~V) obtained after annealing in reducing atmosphere (\ce{N2}) at 300 $^\circ$C, following the same annealing procedure as in Figure~\ref{fig:annealing}. Bright features in the cAFM image correspond to neutral domain walls, showing that their conductance is substantially higher than in the $+P$ and $-P$ domains after the exposure to the reducing environmental conditions.

The conductance map in Figure~\ref{fig:revers}(b) presents the transport properties after the additional annealing step involving synthetic air. We find that the sample exhibits a pronounced contrast between $+P$ and $-P$ ($V_{bias}=4$~V), whereas no specific cAFM signal is measured from the domain walls, which is qualitatively the same behavior as in the as-grown sample shown in Figure~\ref{fig:annealing}(a). This observation leads us to the conclusion, that the changes in domain wall conductance induced by the exposure to reducing atmospheres are reversible and that the initial conditions can be restored by applying adequate annealing procedures. As shown by Figure~\ref{fig:revers}(a) to (d), this process is repeatable, leading to an activation and deactivation of the enhanced transport behavior at the neutral domain walls, which is a key characteristic for the development of domain-wall-based sensor technology.

The same changes in domain wall conductance can also be achieved by alternating annealing in nitrogen, which triggers the onset of conductance, and pure oxygen, which inhibits the conductance. This supports that the changes in oxygen off-stoichiometry are the driving mechanism for the changes in domain wall conductance (see Supplementary Information for more information).


To quantify the changes in domain wall conductance, we analyze local current profiles extracted from the data in Figure~\ref{fig:revers}(a)-(d). The profiles are displayed in Figure~\ref{fig:revers}(e) and show that the current at the neutral domain walls in Figure~\ref{fig:revers}(a) and (c) is about five to six times higher than for the $-P$ domains, whereas a slightly reduced current signal is observed for the walls in Figure~\ref{fig:revers}(b) and (d). Figure~\ref{fig:revers}(f) displays this change in relative domain wall conductance, $\Delta I$, for the consecutive annealing steps, going back and forth between conductive $\left(\Delta I>0\right)$ and insulating $\left(\Delta I<0\right)$ domain walls (see Methods Section for details on the data analysis). The data thus establishes a one-to-one correlation between the atmospheric conditions the sample was exposed to and the domain wall conductance, translating environmental changes into measurable conductance changes. Here, we define the onset of enhanced domain wall conductance as the "sensing" step and its suppression as "reset", inspired by the functional behavior required for the realization of a domain-wall-based oxygen sensor, which is triggered (switches from insulating to conducting) when the environmental oxygen level falls below a certain threshold value, as we emulated by the exposure to reducing conditions.  



\section{Microscopic origin}

DFT calculations presented in previous works have shown that the emergence of p-type conductance at neutral domain walls in hexagonal manganites correlates with the local density of oxygen interstitials.\cite{Schultheiss2020,Skjærvø2018,Schaab2018,Skjærvø2016} Within this picture, there are two basic effects that can explain the observed annealing-induced enhancement of the domain wall conductance (see Figure~\ref{fig:revers}), i.e., (i) a more pronounced loss of oxygen interstitials in $+P$ and $-P$ domains compared to the domain walls during \ce{N_2}-annealing or (ii) a more efficient reoxidation at the domain wall after annealing in \ce{N_2}. Consistent with the reduced defect formation energy for oxygen interstitials at the neutral domain walls, both effects lead to an increase in oxygen concentration relative to the domains and, hence, can contribute to the observed enhancement of the local electronic conduction.\cite{Schultheiss2020,Skjærvø2018,Schaab2018}

\begin{figure}[h]
    \centering
    \includegraphics[width = 0.48\textwidth]{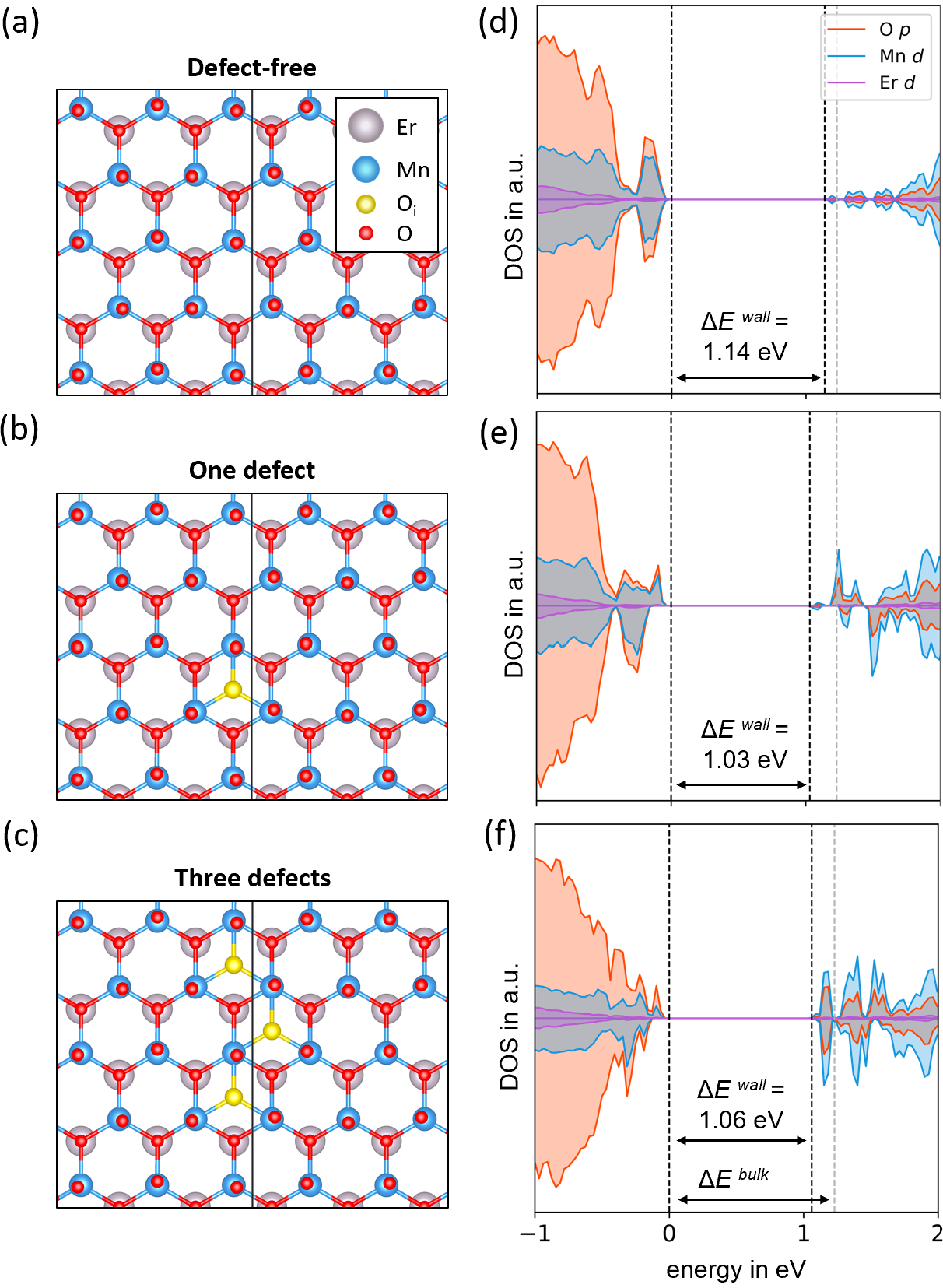}
    \caption{\textbf{Relation between defect density and band structure at domain walls.} (a)-(c) Schematics of the crystal structure of \ce{ErMnO3} with (a) none, (b) one and (c) three oxygen interstitials at the domain wall within our simulation cell. (d)-(f) Density of states calculated for the scenarios illustrated in (a)-(c). Dashed lines represent the Fermi level and the bottom of the conduction band in pristine \ce{ErMnO_3} and $\Delta E^{wall}$ denotes the band gap. The defects illustrated in (c) are distributed over two layers for the calculations.}
    \label{fig:dft}
\end{figure}


The question that remains to be answered is how such changes in oxygen concentration impact the electronic structure, clarifying the microscopic mechanism for enhanced domain wall conductance. To gain insight into the driving mechanism for the enhanced conductance at the neutral domain walls, we perform DFT calculations for different concentrations of oxygen interstitials at domain walls in \ce{ErMnO_3} and determine the density of states (DOS) as presented in Figure~\ref{fig:dft}. \ce{ErMnO_3} consists of alternating Er layers and Mn-O layers with corner-sharing trigonal bipyramids of \ce{Mn^3+} and \ce{O^2-}. The energetically most favorable position for oxygen interstitials is at one of the six equivalent lattice sites between the \ce{Mn} atoms in the \ce{Mn}-\ce{O} planes\cite{Skjærvø2016} (see Supplementary Information for details of the DFT calculations).

Figures~\ref{fig:dft}(a) and (d) show the model of a neutral domain wall and the calculated DOS. Consistent with literature\cite{Skjærvø2016, Skjærvø2018, Schaab2018}, we observe qualitatively the same electronic structure as in the bulk with only a subtly smaller band gap (i.e., $\Delta E^{bulk}=1.23$~eV and $\Delta E^{wall}=1.14$~eV).
When introducing one oxygen interstitial at the domain wall, corresponding to a local off-stoichiometry of $\delta = 0.08$ at the domain wall (where $\delta$ is defined in terms of the volume encompassed by the domain wall width), a localized and non-bonding defect state arises at the bottom of the conduction band, with the corresponding bonding states situated at the bottom of the valence band (Figure~\ref{fig:dft}(b) and (e)). This effect is caused by a small shift of the oxygen interstitial away from its central position, changing the valence state of two Mn atoms from \ce{Mn^3+} to \ce{Mn^4+}. This implies that the electronic transport at the neutral domain walls occurs through a p-type polaron hopping mechanism \cite{Skjærvø2016}. We also observe a lowering of the band gap to $\Delta E^{wall}=1.03$~eV, which is \SI{0.11}{eV} lower than for a defect-free domain wall, while a single defect in bulk leads to a considerably smaller band gap of $E^{wall}=0.81$~eV. This can explain the observed disparity in the relative bulk-wall conductance after annealing in a reducing atmosphere versus synthetic air, as the bulk has an additional effective charge carrier contribution from the band gap reduction even at lower defect concentrations. As the concentration of oxygen interstitials at the neutral domain wall increases to $\delta=0.24$ in Figure~\ref{fig:dft}(c) and (f), the defect states become less localized and the band gap remains close to the value for one defect.


A polaron hopping mechanism is inferred at the domain wall from the DFT results. The charge carrier density modulation, caused by holes charge compensating oxygen interstitials, is identified as the main reason for the enhanced domain wall conductance in the scans shown in Figures~\ref{fig:annealing} and~\ref{fig:revers}. While the presence of interstitials at domain walls also results in a smaller local band gap and minor changes to the electron/hole mobility, these effects are too small to account for the experimental observations in Figure~\ref{fig:revers}.

\section{Outlook}

\begin{figure}
	\centering
	\includegraphics[width = 0.48\textwidth]{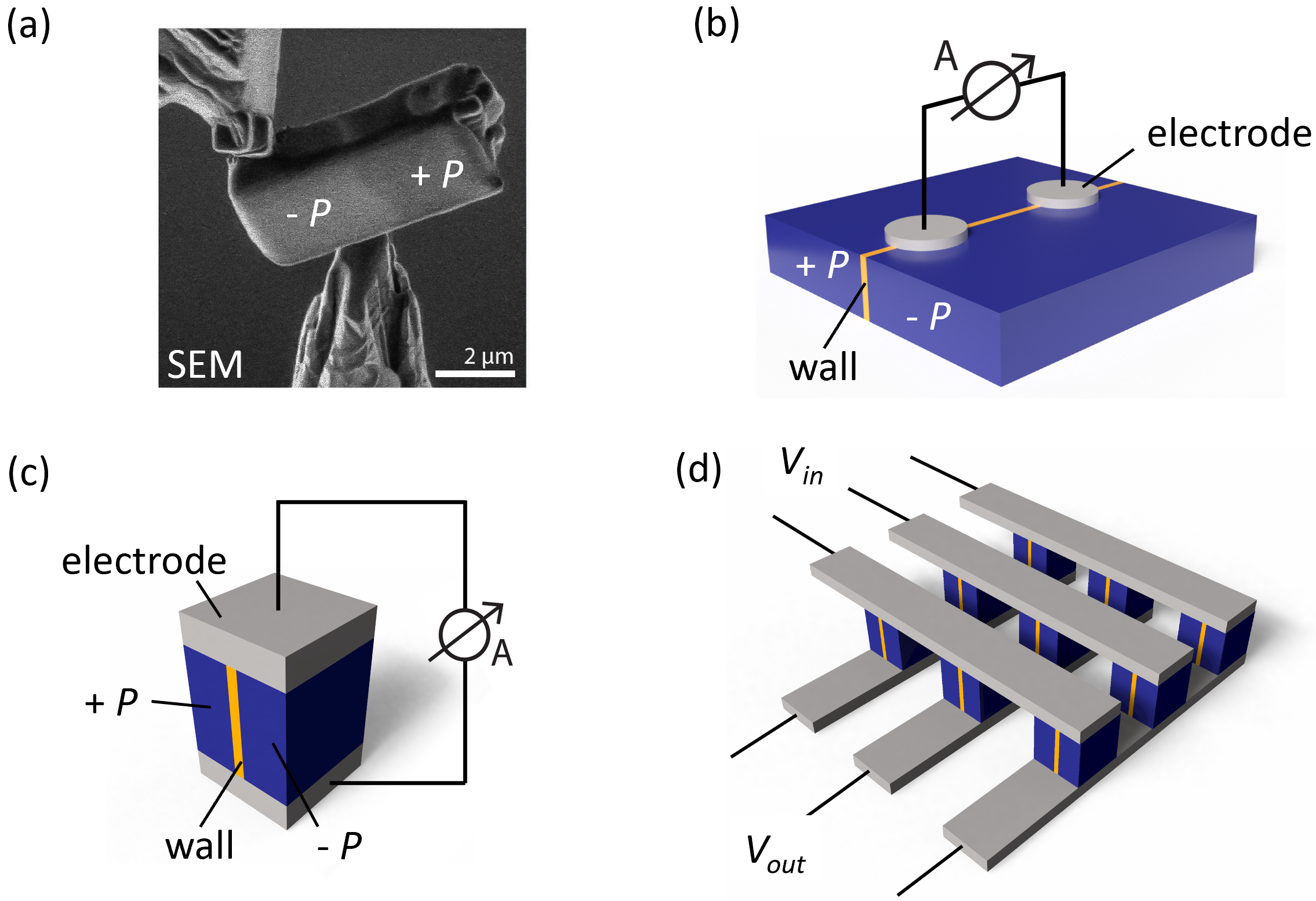}
	\caption{\textbf{Concept for domain wall-based sensors.} (a) SEM image of a volume with a single ferroelectric domain wall extracted from a bulk sample using FIB, demonstrating the general possibility to isolate individual walls for sensor development. (b),(c) Illustration of different sensing geometries based on a single ferroelectric domain wall using lateral (b) or transversal (c) electrode arrangements. (d) Crossbar geometry combining multiple transversal electrode arrangements into an array for spatially resolved sensing.}
	\label{fig:concept}
\end{figure}

The results presented in Figure~\ref{fig:annealing} to~\ref{fig:dft} show a direct relation between environmental conditions (here, the concentration of oxygen in the atmosphere) and the electronic conduction at neutral ferroelectric domain walls in \ce{Er(Mn{,}Ti)O}$_3$. All experiments were conducted on domain wall networks at the surface of millimeter-sized single crystals. While it is clear that the physical and chemical domain wall properties will not change qualitatively upon down-scaling\cite{Mosberg2019}, it remains to be demonstrated that smaller volumes can be prepared without losing the domain walls. To test the general feasibility, we use scanning electron microscopy (SEM) in combination with a focused ion beam (FIB) to extract a volume of $2 \times 2 \times 8$~\SI{}{\micro m^3} with an individual domain wall as displayed in Figure~\ref{fig:concept}(a). The SEM image shows the isolated specimen with one domain wall in the center, separating two ferroelectric domains (bright and dark) of opposite polarization.

On the one hand, the successful extraction of a single domain wall demonstrates the possibility to scale domain-wall-based sensors and to work with individual walls. On the other hand, it shows the opportunity to work in different geometries as illustrated in Figure~\ref{fig:concept}(b) and (c). For example, electrodes can be patterned on the sample surface to measure environmentally driven changes in domain wall resistivity as outlined in Figure~\ref{fig:concept}(b). Alternatively, micrometer-sized sensor units may be realized as sketched in Figure~\ref{fig:concept}(c), which may be used individually or in crossbar arrangements (Figure~\ref{fig:concept}(d)), enabling spatially resolved sensing. Independent of the specific geometry, the domain walls represent the active sensing medium, switching from insulating to conducting (or vice versa) as function of the environmental conditions, as demonstrated in Figure~\ref{fig:revers}.


In conclusion, the results presented in this work introduce ferroelectric domain walls as promising active elements for environmental sensors. We demonstrated that by exposing neutral domain walls in \ce{Er(Mn{,}Ti)O}$_3$ to reducing and oxidizing atmospheres, their conductance can be reversibly switched between insulating and conducting behavior, translating changes in atmospheric conditions into electronic signals. The effect originates from the distinct interaction of oxygen defects with the domain walls, which can readily be expanded towards other oxide systems.~\cite{Campanini2020,Rojac2017, Eggestad2024} Our concept for domain-wall-based environmental sensors gives an additional dimension to the field of domain wall nanoelectronics and expands the pool of functional materials for sensor technology towards ferroelectric domain walls.



\section{Methods}
\subsection{Synthesis and sample preparation}
\ce{Er(Mn,Ti)O_3} single crystals are grown by the pressurized floating zone method.\cite{Yan2015} The samples are then oriented by Laue diffraction and cut such that the $c$-axis is perpendicular to the sample surface ([001]-oriented), resulting in an out-of-plane polarization where the domain polarization can be either pointing out of the sample surface ($+P$, up) or into the sample surface ($-P$, down). The crystals are lapped with a \SI{9}{\micro\m}-grained \ce{Al_2O_3} water suspension and polished with a silica slurry, resulting in a surface roughness of approximately \SI{0.5}{nm}.

\subsection{Microscopy}
The cAFM scans are collected using a commercial atomic force microscope (Asylumn Research, Cypher ES Environmental AFM). Diamond-coated tips are used and the cAFM maps are conducted with the positive bias voltage $V_{bias}$ applied to the back electrode while the tip is grounded. FIB cutting and SEM imaging was performed using a Thermo Fisher Scientific G4UX Dual-beam FIB-SEM.

\subsection{Annealing and heating}
The annealing is performed in an Entech Tube furnace with a cooling/heating rate of 200 $^\circ$C/h and a maximum temperature of 300 $^\circ$C. The dwell time at the maximum temperature is 48 hrs. Before the annealing, the furnace is evacuated three times below 0.1 mbar to ensure a pure annealing atmosphere. A continuous gas flow is maintained during all steps of the annealing.

Additional heating experiments are performed inside the Cypher ES Environmental AFM. The sample chamber is first flushed with the desired gas and then an overpressure of $\approx\SI{300}{mbar}$ is built up to avoid contaminants from the outside. The detailed heating procedures can be found in the Supplementary Information.

\subsection{Data analysis}

The relative domain wall conductance shown in Figure~\ref{fig:revers}(f) is obtained from the profiles in Figure~\ref{fig:revers}(e). The profiles are averaged over a width of \SI{290}{\micro m}, to reduce the influence of noise in the scan data. To make the profiles comparable, the current values are normalized such that the darker domain has an average current of 0 and the brighter domain of 1. For this, all current values are transformed as $I^\prime = \frac{I-I_{dark}}{I_{bright}-I_{dark}}$.
For deriving the normalized domain wall current, the curves are fitted with a function of the form
\begin{equation*}
    I(x) = \underbrace{\frac{A_s}{1+e^{-2\cdot \sigma_s\cdot(x-x_0^s)}}}_{I_{step}(x)} + \underbrace{A_g\cdot e^{-\frac{(x-x_0^g)^2}{\sigma_g^2}}}_{I_{peak}(x)}+C
\end{equation*}
where $I_{step}$ is representing the difference in conductance in between the two domain states, while $I_{peak}$ represents the enhanced or reduced conductance at the domain wall. The distance along the profile line is represented as $x$. From this fit, the height of the additional current change at the domain wall $\Delta I=A_g$ can be derived for all four profiles.

\subsection{Density Functional Theory}
    
DFT calculations were done using the VASP \cite{Kresse1996, Kresse1999} code with the PBEsol functional \cite{Perdew2008} and GGA+$U$ \cite{Dudarev1998}, to which a Hubbard $U$ of \SI{4}{eV} was applied to the \ce{Mn} 3$d$ orbitals. The PAW method \cite{Blochl1994} with a cutoff energy of \SI{550}{eV} was used together with the Er\_3, Mn\_pv and O pseudopotentials supplied with VASP. A $\varGamma$-centered $4\times 4\times 2$ k-point mesh was used for the 30 atom unit cell, with similar densities for larger supercells. For bulk calculations, atomic positions were relaxed until the residual forces were below  \SI{0.001}{eV/ \angstrom} for the 30 atom cell, and \SI{0.005}{eV/ \angstrom} for the $2 \times 2 \times 1$ supercell. For neutral domain walls, a $1 \times 6 \times 1$ supercell was used for the pristine wall and a $2 \times 6 \times 1$ supercell with defects present. The geometry was  optimized until the forces on the atoms were below \SI{0.02}{eV/ \angstrom}. The non-collinear magnetic order on the \ce{Mn} atoms was approximated by a frustrated collinear antiferromagnetic order \cite{Medvedeva_2000} and kept continuous across the domain walls (see Supplementary Information for details on magnetic order in defect calculations). Band structure unfolding was done using the easyunfold package \cite{Zhu2024}. 
 
\begin{acknowledgments}
The authors thank Kristoffer Eggestad, Benjamin A.D. Williamson, Xuejian Wang, and Frank Wendler for fruitful discussions and valuable input. D.M. acknowledges NTNU for support through the Onsager Fellowship Program and the Outstanding Academic Fellow Program. D.M., L.R., U.L., and J.H. acknowledge funding from the European Research Council (ERC) under the European Union's Horizon 2020 Research and Innovation Program (Grant Agreement No. 863691). The Research Council of Norway is acknowledged for the support to the Norwegian Micro- and Nano-Fabrication Facility, NorFab, Project No. 295864 and the Norwegian Center for Transmission Electron Microscopy, NORTEM (No. 197405).
I.C.S. and S.M.S. acknowledge computational resources for density functional theory simulations provided by Sigma2 - the National Infrastructure for High Performance Computing and Data Storage in Norway through project NN9264K. I.C.S. and S.M.S. also acknowledge support by the Research Council of Norway through project 302506. J.S. acknowledges the support of the Alexander von Humboldt Foundation through a Feodor-Lynen research fellowship.
\end{acknowledgments}

\section*{Data Availability Statement}

The data that support the findings of this study are available from the corresponding author upon reasonable request.

\bibliography{annealing_bib}

\end{document}



\title[Supplementary Information: Ferroelectric domain walls for environmental sensors]{Supplementary Information: Ferroelectric domain walls for environmental sensors}
\author{L. Richarz}
\affiliation{ 
Department of Materials Science and Engineering, NTNU Norwegian University of Science and Technology, Trondheim, Norway
}%

\author{I. C. Skogvoll}
\author{E. Y. Tokle}
\affiliation{ 
Department of Materials Science and Engineering, NTNU Norwegian University of Science and Technology, Trondheim, Norway
}%
\author{K. A. Hunnestad}
\affiliation{ 
Department of Materials Science and Engineering, NTNU Norwegian University of Science and Technology, Trondheim, Norway
}%
\affiliation{ 
Department of Electronic Systems, NTNU Norwegian University of Science and Technology, Trondheim, Norway
}%
\author{U. Ludacka}
\affiliation{ 
Department of Materials Science and Engineering, NTNU Norwegian University of Science and Technology, Trondheim, Norway
}%
\affiliation{%
Department of Physics, NTNU Norwegian University of Science and Technology, Trondheim, Norway
}%

\author{J. He}
\affiliation{ 
Department of Materials Science and Engineering, NTNU Norwegian University of Science and Technology, Trondheim, Norway
}%

\author{E. Bourret}
\affiliation{%
Materials Sciences Division, Lawrence Berkeley National Laboratory, Berkeley, CA, USA
}%

\author{Z. Yan}
\affiliation{%
Materials Sciences Division, Lawrence Berkeley National Laboratory, Berkeley, CA, USA
}%
\affiliation{%
Department of Physics, ETH Zurich, Zurich, Switzerland
}%

\author{A.T.J. van Helvoort}
\affiliation{%
Department of Physics, NTNU Norwegian University of Science and Technology, Trondheim, Norway
}%

\author{J. Schultheiß}
\affiliation{
Department of Materials Science and Engineering, NTNU Norwegian University of Science and Technology, Trondheim, Norway
}%

\author{S. M. Selbach}
\affiliation{ 
Department of Materials Science and Engineering, NTNU Norwegian University of Science and Technology, Trondheim, Norway
}%

\author{D. Meier}
\affiliation{ 
Department of Materials Science and Engineering, NTNU Norwegian University of Science and Technology, Trondheim, Norway
}%


\maketitle
\beginsupplement

\section*{Reversibility in \ce{N_2}-\ce{O_2}-\ce{N_2}-Annealing}

\begin{figure}[h]
	\centering
	\includegraphics[width=0.8\textwidth]{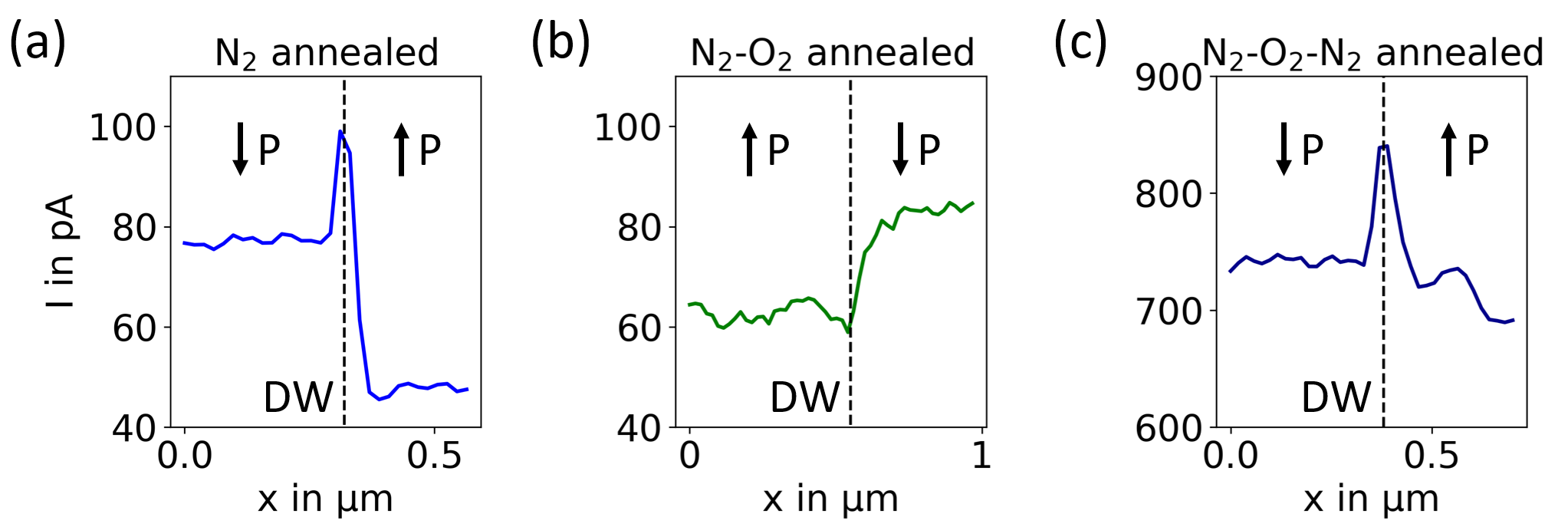}
	\caption{Profiles across a domain wall (approximate position marked by the black dashed line) after different steps of the annealing cycle. (a) After the sample was annealed in \ce{N_2} a clear increase in conductance is visible at the domain wall. (b) After subsequent \ce{O_2} annealing no enhanced conductance at the wall can be observed. (c) After the sample is annealed in \ce{N_2} again, the domain wall shows increased conductance. The corresponding cAFM scans can be found in Figure~\ref{fig:o2}.}
	\label{fig:o2_dws}
\end{figure}

To see if a transition from conducting to insulating domain walls (shown in Figure~2 in the main text) can also be achieved by pure oxygen annealing, we perform additional annealing experiments. An \ce{N_2}-annealed sample is further annealed in \ce{O_2} at \SI{300}{\degreeCelsius} for \SI{48}{hrs} with \SI{200}{\degreeCelsius/h} cooling and heating rate. Profiles across a domain wall after annealing in \ce{N_2} and after the subsequent \ce{O_2} annealing are shown in Figure~\ref{fig:o2_dws}(a) and (b) respectively. As visible, the conductance is clearly enhanced after \ce{N_2} annealing. After the following \ce{O_2} annealing, no enhanced domain wall conductance is visible any more. This shows that the transition from conducting to insulating domain walls by in-situ heating that was shown in Figure~2 in the main text can also be achieved by annealing in \ce{O_2}. Subsequent annealing in \ce{N_2} restores the enhanced conductance at the domain walls (see Figure~S2), again demonstrating the reversibility of the process.

\begin{figure}[h]
	\centering
	\includegraphics[width=\textwidth]{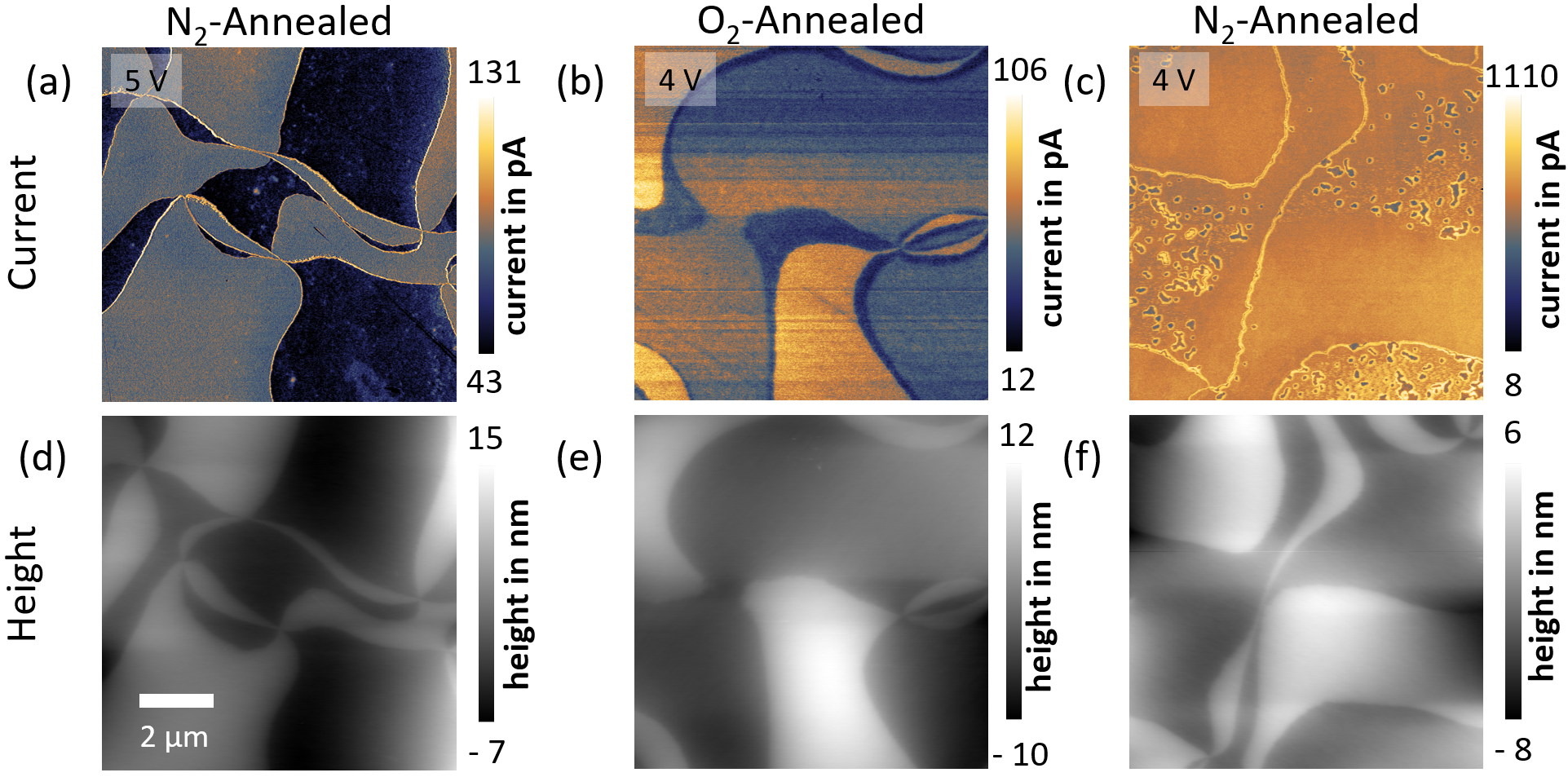}
	\caption{Comparison of the domains visible in the current (a)-(c) and the height channel (d)-(f) at different points in the annealing cycle. (a) and (d) are recorded after the sample is annealed in \ce{N_2} for the first time. (b) and (e) are recorded after the same sample is subsequently annealed in \ce{O_2}. Data (c) and (f) are recorded after a subsequent second annealing in \ce{N_2}. All images are recorded with a grounded, diamond-coated DEP$01$ tip and a voltage of (a) \SI{5}{V}, (b) and (c) \SI{4}{V} applied to the back electrode. Current and height are always recorded simultaneously but the positions on the sample are different for the different annealing steps to avoid scan imprints.}
	\label{fig:o2}
\end{figure}

This annealing cycle (\ce{N_2}-\ce{O_2}-\ce{N_2}) is, however, accompanied  by significant switching of the surface domains, as visible when comparing the current data in the corresponding cAFM scans presented in  Figure~\ref{fig:o2}(b) and (c) with the respective topography in Figure~\ref{fig:o2}(e) and (f). The topography signal carries the information about the domain state in the as-grown state due to selective etching during the sample preparation~\cite{Safrankova1967} and can thus be used as a reference for changes in the domain configuration. The switching behavior can be explained by a gradient in oxygen off-stoichiometry, where the annealing affects only the surface-near region.~\cite{Wang2015} 

The initial \ce{N_2} annealing results in a slight expansion of the bright $-P$ domains (see Figure~\ref{fig:o2}(a)). This effect is consistent with a downwards pointing electric field, induced by a reduction in the amount of negatively charged interstitials or introduction of positively charged vacancies at the surface. This is the expected case after annealing in reducing atmospheres. This slight variation in the domain size can be observed on most \ce{N_2} annealed samples, and the flexibility in the domain wall position would need to be taken into account for a potential sensor device. The subsequent \ce{O_2} annealing then leads to an expansion of the darker $+P$ domains. Again, this is consistent with the expected increase in the amount of negatively charged oxygen interstitials at the surface, introducing an upwards pointing electric field close to the sample surface. The final \ce{N_2} annealing results in pronounced switching, where the dark $+P$ domains are contracted to thin lines and bubbles. Such a strong amount of switching has not been previously observed in the samples that were re-oxidized by in-situ heating (see Figure~2 in the main text for reference), indicating that oxidation and reduction during annealing of the samples have different dynamics. This highlights the importance of carefully controlling the annealing parameters if full reversibility is desired.

\section*{Annealing History}

As discussed in the last section, the reversible switching of the conductance at neutral domain walls is a non-trivial process, as the gradient of the oxygen off-stoichiometry has to be precisely controlled to avoid large-scale switching of the surface domains. 

Table~\ref{tab:annealing_history} describes the full heating history of the sample used for collecting the scans shown in Figure~2. The scans in Figure~2 are taken before (a) and after (b) the heating on  day 9, as well as before the heating on day 45 (c) and after the heating on day 50 (d).

\begin{table}[h!]
	\centering
	\resizebox{\textwidth}{!}{
        \begin{tabular}{r|ccccc|l}
    		Day & Atmosphere & $T_{max}$ in \SI{}{\degreeCelsius} & Time in h & Rate in \SI{}{\degreeCelsius/h} & Instrument & Domain walls \\
    		\hline
    		1 &	\ce{N_2} &	300 &	48 &	200	& Entech Tube Furnace &	conducting\\
    		3 &	\ce{N_2} &	200	& 2	& 1800 &	Cypher AFM &	conducting\\
    		9 &	\ce{N_2} \& Air &	200	& 2	& stepwise &	Cypher AFM& 	not conducting\\
    		15 &	\ce{N_2} &	250	& 2	& 1800& 	Cypher AFM &	not conducting\\
    		28 &	\ce{N_2} &	300	& 48 &	200 &	Entech Tube Furnace &	conducting\\
    		43 &	\ce{N_2} &	100	& 1 &	stepwise &	Cypher AFM &	conducting\\
    		45 &	\ce{N_2} &	200	& 0.5 &	stepwise &	Cypher AFM &	conducting\\
    		50 &	\ce{N_2} \& Air &	200	& 1	& stepwise &	Cypher AFM &	not conducting\\
    	\end{tabular}
     }
\caption{Annealing history of the sample shown in Figure~2. $T_{max}$ describes the maximum temperature that was reached during the heating, the time gives the approximate time the sample spent at that maximum temperature, and the rate gives the cooling and heating rate. Many of the heating experiments in the Cypher AFM were performed stepwise, and no absolute heating rate can be given. The last column describes the relative domain wall conductance after the respective heating cycle.}
\label{tab:annealing_history}
\end{table}

\section*{Density functional theory}
\begin{figure}[h]
    \centering
    \includegraphics[width=0.7\textwidth]{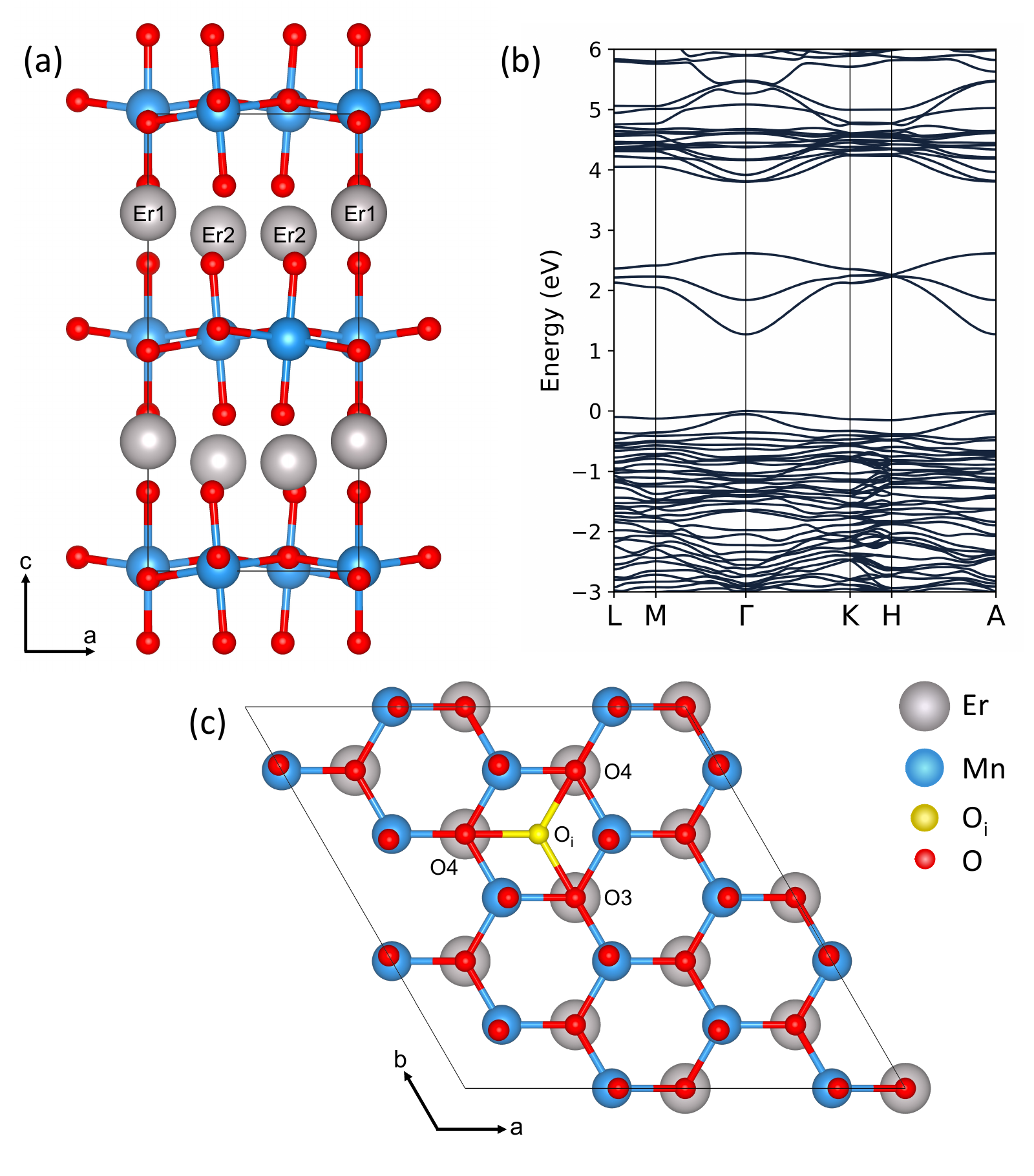}
    \caption{(a) The unit cell structure of \ce{ErMnO3}, with \ce{Mn} atoms indicated in blue, \ce{O} in red and \ce{Er} in grey. (b) The band structure of \ce{ErMnO3}. The calculated band gap is $E_g = \SI{1.23}{eV}$. (c) (001) plane of the unrelaxed $2\times 2 \times 1$ supercell, with $\text{O}_\text{i}$ in a stable position (yellow).}
    \label{fig:structure}
\end{figure}

The structure of \ce{ErMnO3} consists of  alternating \ce{Er} and \ce{Mn}-\ce{O} layers in the $ab$-plane, with the latter forming corner-sharing trigonal bipyramids of \ce{Mn^3+} and \ce{O}. Within each layer, the two \ce{Er} atoms in the Er2 position are displaced down while the single Er1 is displaced upwards in the $c$-direction, giving rise to the polarization, see Figure~\ref{fig:structure}. The largest dispersion in the band structure of \ce{ErMnO3} is observed in high-symmetry directions corresponding to intra-layer directions within the unit cell. The calculated band gap is $E_g = \SI{1.23}{eV}$. 

The most energetically stable position for interstitial oxygen ($\text{O}_\text{i}$) in the lattice is within the \ce{Mn-O} layers, centered between the three \ce{Mn} atoms, such that there is six equivalent lattice sites within one unit cell, as shown in Figure~\ref{fig:structure}(c). In this position, $\text{O}_\text{i}$ experiences a triple well potential where each minima is shifted slightly from the central position towards two \ce{Mn} atoms. These are then oxidized from \ce{Mn^3+} to \ce{Mn^4+} through a partial charge transfer, resulting in two \ce{Mn-O} bond lengths of equal size and one elongated bond to the latter \ce{Mn} atom. As shown in previous work on $\text{O}_\text{i}$ in \ce{YMnO3}\cite{Skjærvø2016}, this triple well potential is asymmetric as there is an additional energy gain for $\text{O}_\text{i}$ to displace towards the O3 position, which is the trimerization center of the corner-sharing \ce{MnO5} trigonal bipyramids. 
\begin{figure}[h]
    \centering
    \includegraphics[width=0.8\textwidth]{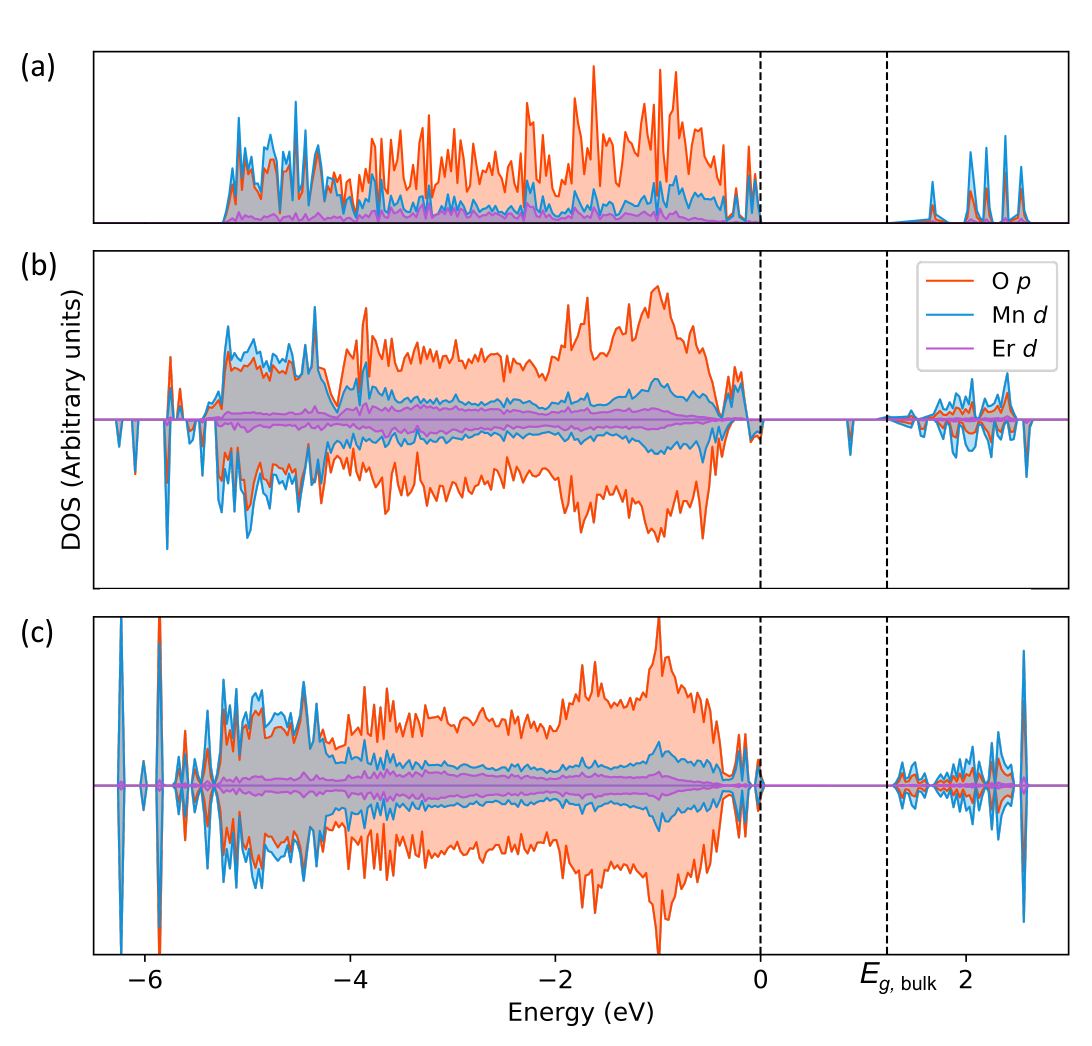}
    \caption{Density of states calculated for the $2\times 2 \times 1$ supercell of \ce{ErMnO3} with (a) no $\text{O}_\text{i}$, (b) one $\text{O}_\text{i}$ and (c) two $\text{O}_\text{i}$. Upper and lower plots indicate up and down spin channels, respectively, except for the case of no interstitials, as we then have perfect antiferromagnetic order. Dashed vertical lines indicate (for this plot and all following density of states) the Fermi level and bottom of the conduction band for bulk \ce{ErMnO3}.}
    \label{fig:DFT_defects}
\end{figure}

\begin{figure}[h]
    \centering
    \includegraphics[width=\textwidth]{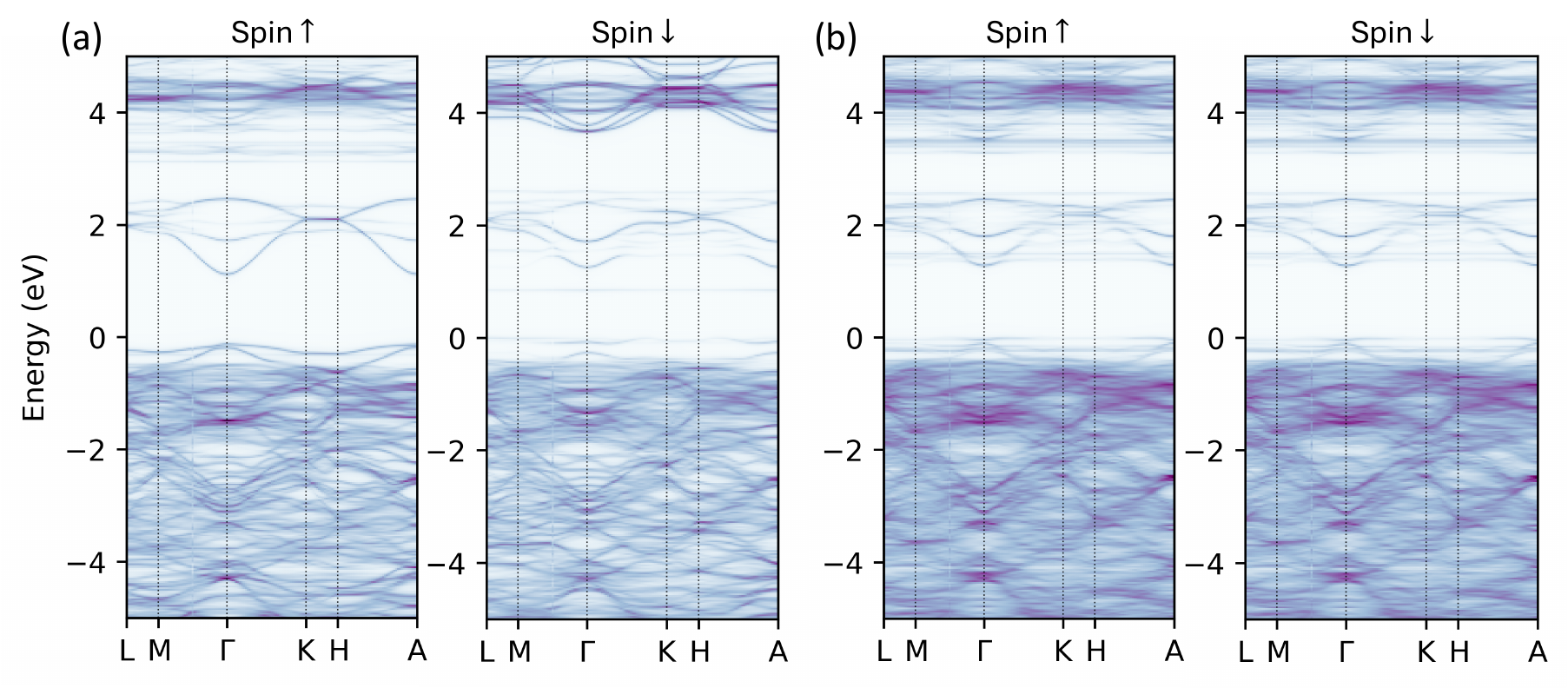}
    \caption{Unfolded band structures for the $2\times 2 \times 1$ supercell, with (a) one $\text{O}_\text{i}$ and (b) two $\text{O}_\text{i}$, separated into spin up and down channels as indicated above the plot. Defect levels appear as localized horizontal bands within and below the counduction band.}
    \label{fig:DFT_defects_band}
\end{figure}

The change in electronic structure from an increase in interstitial oxygen in the lattice was calculated using a $2\times 2\times 1$ supercell with a single $\text{O}_\text{i}$ in a stable position, corresponding to an off-stoichiometry of $\delta = 0.04$. As displayed in Figure~\ref{fig:DFT_defects}, we see the emergence of a localized and non-bonding defect state within the band gap, with the bonding states situated below the valence band. The band gap is reduced to $E_g = \SI{0.81}{eV}$. For $\delta = 0.08$, which is achieved by a single oxygen defect in each \ce{Mn-O} layer, the localized nature of the defect states is diminished and they move into the bottom of the conduction band. The asymmetry in the defect states within each spin channel is also no longer present. The unfolded band structures for $\delta = 0.04$ and $\delta = 0.08$ are presented in Figure~\ref{fig:DFT_defects_band}. 

\begin{figure}[h]
    \centering
    \includegraphics[width=0.8\textwidth]{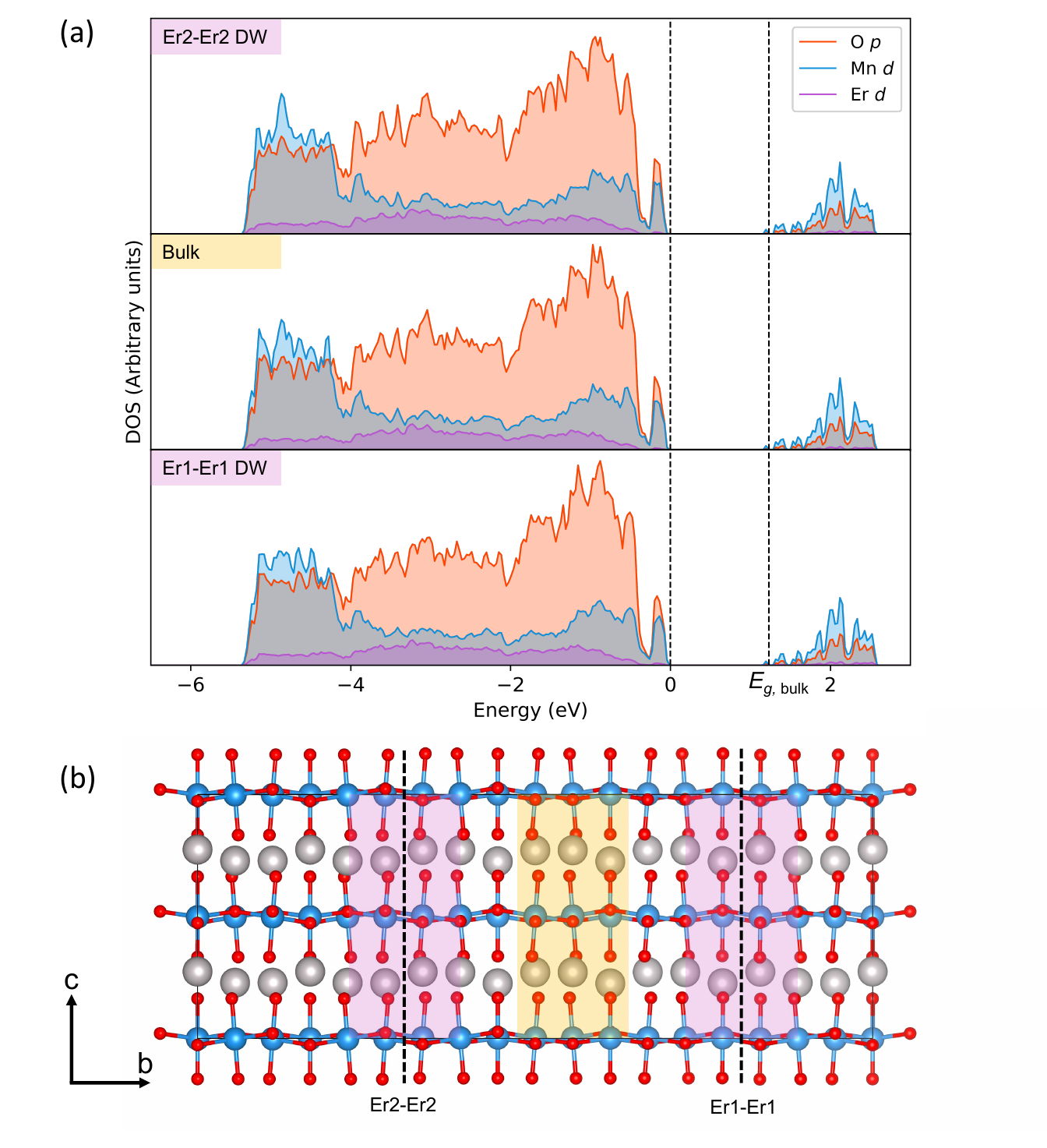}
    \caption{ (a) Density of states calculated for a $1\times 6 \times 1$ supercell with Er2-Er2 and Er1-Er1 terminated neutral domain walls. (b) Relaxed geometry of the supercell as seen from the (100) plane, with the position of each domain wall marked by vertical dashed lines. The three plots show the layer-resolved density of states for a unit cell centered at the position of each domain wall, as well as in between each wall (in domain), assigned as bulk. The position of each layer and domain wall type is indicated by the corresponding color displayed in the structure below.}
    \label{fig:DFT_DW}
\end{figure}

For the neutral domain wall calculations, a $1\times 6 \times 1$ cell was used for the pristine domain wall, while a supercell of dimensions $2\times 6 \times 1$ was used when simulating $\text{O}_\text{i}$ adjacent to the wall to minimize defect-defect interactions. The symmetry alteration introduced by a domain wall is expected to give rise to a locally modified electronic structure as opposed to the bulk domains. However, since the polarization is parallel to a close to atomically sharp domain wall, there are no bound charges. As a result, the density of states and electronic structure is qualitatively similar to the bulk, albeit with a slightly lower band gap of $E_g = \SI{1.14}{eV}$, see Figure~\ref{fig:DFT_DW}. 

\begin{figure}[h]
    \centering
    \includegraphics[width=0.7\textwidth]{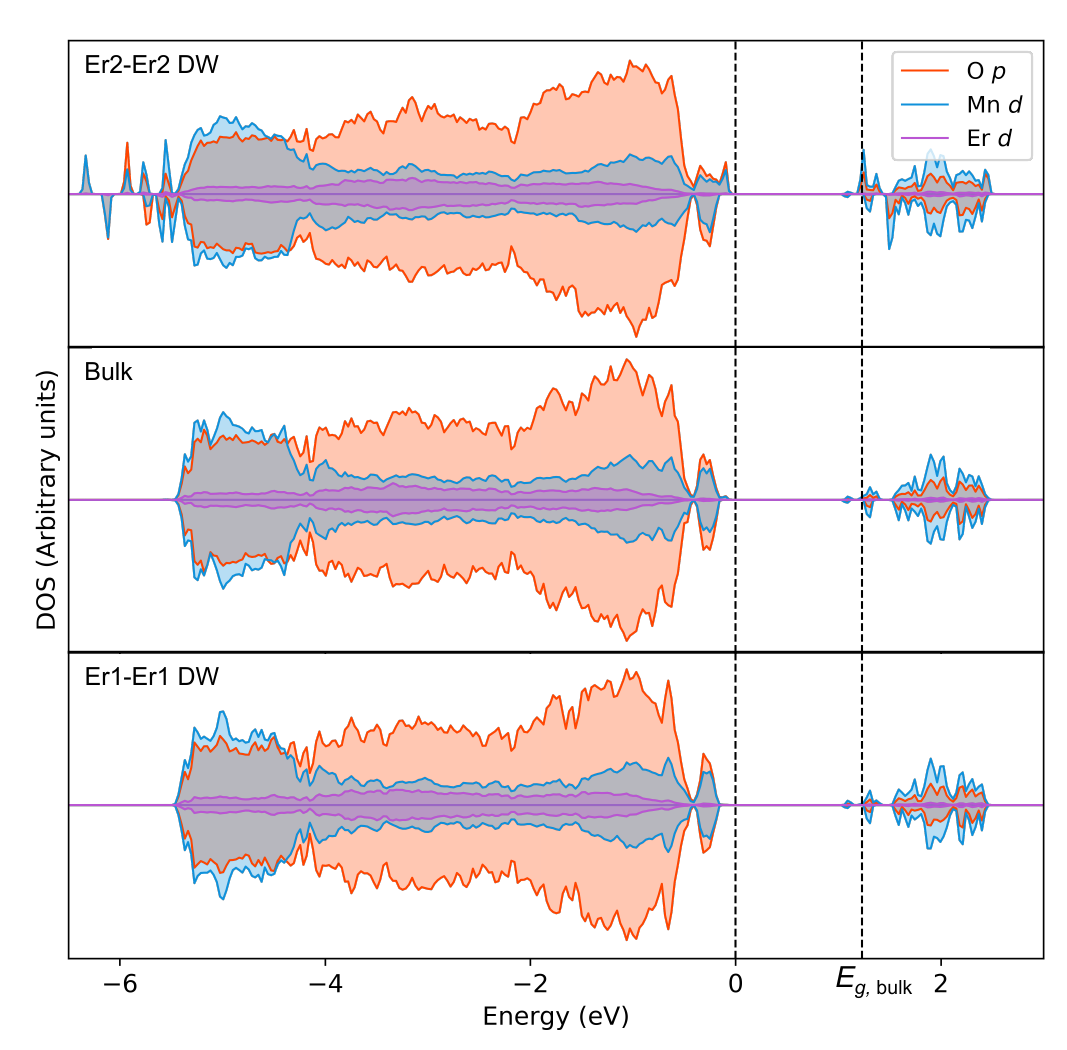}
    \caption{Layer-projected density of states calculated for a $2\times 6 \times 1$ supercell, with a single $\text{O}_\text{i}$ positioned adjacent to the Er2-Er2 terminated domain wall. The positions of the layers follow that of Figure~\ref{fig:DFT_DW}. The defect levels attributed to the interstitials can be seen below the valence band and in the lower parts of the conduction band, in the Er2-Er2 domain wall layer.}
    \label{fig:DFT_DW_defect}
\end{figure}

\begin{figure}[h]
    \centering
    \includegraphics[width=0.7\textwidth]{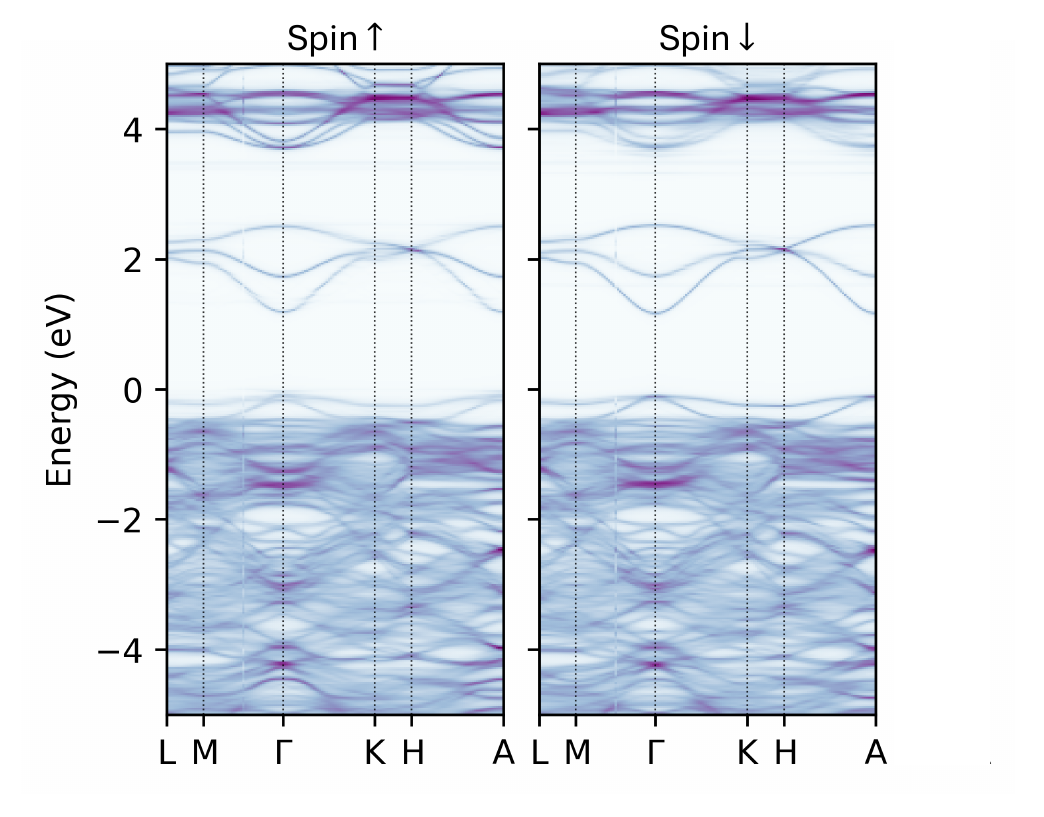}
    \caption{Unfolded band structure for spin up and down channels for the entire $2\times 6 \times 1$ supercell, with a single oxygen interstitial adjacent to the Er2-Er2 domain wall. The localized defect states can be seen as horizontal lines within the conduction band. The low visibility of the lines results from the size of the supercell.}
    \label{fig:DFT_DW_defect_band}
\end{figure}

For simulating defects adjacent to the domain wall, all $\text{O}_\text{i}$ were placed in stable configurations that minimized defect-defect interactions. The layer-projected density of states and unfolded band structures in the case of a single $\text{O}_\text{i}$ adjacent to the neutral domain wall are presented in Figures~\ref{fig:DFT_DW_defect} and \ref{fig:DFT_DW_defect_band}. Both Er2-Er2 and Er1-Er1 domain walls were tested, but showed no significant differences in terms of electronic structure or formation energies, which is why only the Er2-Er2 wall is presented here. The band gap is locally lowered to $E_g = \SI{1.03}{eV}$, and the defect states can be seen in both the bottom of the conduction band and valence band. For three defects adjacent to the neutral domain wall, the defect states become more localized but there is no further lowering of the band gap. We note that the frustrated collinear antiferromagnetic order introduces an artificial variation in the energy and the top and bottom of the valence and conduction bands, respectively, as discussed in the following subsection. 

\subsection*{Frustrated collinear magnetism}

Although \ce{ErMnO3} is paramagnetic at room temperature, the antiferromagnetic ground state is used for DFT calculations to reduce the computational complexity. Furthermore, the non-collinear frustrated antiferromagnetism is approximated by a frustrated collinear antiferromagnetic structure (F-AFM) as this is well known to reproduce the main electronic structure features of rare earth hexagonal manganites\cite{Medvedeva_2000}. However when introducing $\text{O}_\text{i}$ into the crystal structure, some additional considerations need to be made when evaluating the results. 

Firstly, the F-AFM order introduces an additional asymmetry to the triple well potential that the $\text{O}_\text{i}$ experiences in bulk, as for only 1/3 of possible $\text{O}_\text{i}$ sites the adjacent \ce{Mn^4+} atoms have equal spin, the majority spin of that respective \ce{Mn-O} layer. Placing $\text{O}_\text{i}$ in one of the latter 2/3 positions  yields an extra energy gain coming from the electronic repulsion associated with the two \ce{Mn^4+} atoms having opposite spins. This will also result in $\text{O}_\text{i}$ distorting such that the three \ce{Mn-O} bonds are all of different lengths. Thus, for the bulk calculations, defects were placed such that the magnetic order aligned with the most stable configuration possible. 

\begin{figure}[h]
    \centering
    \includegraphics[width=0.7\textwidth]{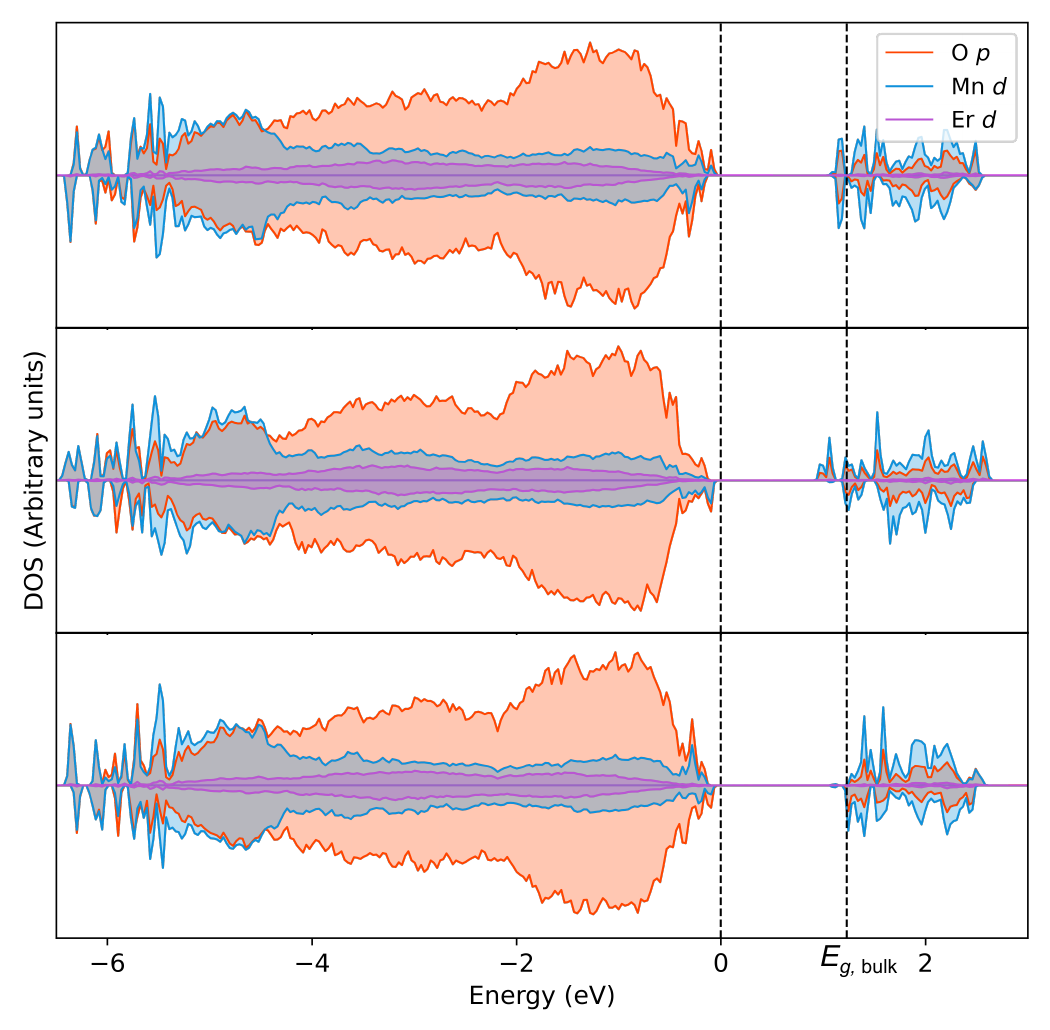}
    \caption{Density of states calculated for a $2\times 6 \times 1$ supercell with three $\text{O}_\text{i}$, projected onto the unit cell centered at the Er2-Er2 domain wall. Each plot shows a potential defect configuration, along with the variation in the location of the defect levels.}
    \label{fig:DFT_DW_d3_configurations}
\end{figure}

In addition to the inequality between interstitial sites stemming from the synthetic F-AFM order, the symmetry-breaking of the domain wall leads to an added artifact in terms of the position of the defect levels and ground state energy. This is because as the crystal structure switches from one domain to the other, while supporting a continuous magnetic order, each potential interstitial site experiences a different local geometry and magnetic surroundings. The relaxed $\text{O}_\text{i}$ is thus always in a distorted structure relative to the most stable bulk position. In the case of one and three $\text{O}_\text{i}$ adjacent to the domain wall, all configurations that were inequivalent in terms of structural and magnetic order were evaluated. The number of possible configurations for three defects are limited by restrictions on the minimum defect-defect distance. A final configuration was chosen, displaying a representative average in terms of the defect energy, as well as the location of the defect levels and thus the band gap. The variation in terms of ground state energy increases with the number of defects, being $\Delta E_{\text{O}_\text{i}} = \SI{0.09}{eV}$ for one $\text{O}_\text{i}$ and $\Delta E_{\text{3O}_\text{i}}  = \SI{0.11}{eV}$ for three $\text{O}_\text{i}$. An example of the variation within the position of the defect levels for three defects is presented in Figure~\ref{fig:DFT_DW_d3_configurations}. This fluctuation in the band gap results from the change in stability between the different configurations due to the domain wall naturally removing the true trimerization points, but also from the artificial magnetic order.

\FloatBarrier
\bibliography{annealing_bib}